\documentclass[11pt,twoside]{article}
\usepackage{asp2010}

\resetcounters

\begin{document}

\title{MIS: a MIRIAD Interferometry Singledish toolkit}
\author{Marc W. Pound$^1$, and Peter Teuben$^1$
\affil{$^1$Astronomy Department, University of Maryland, College Park}}

\begin{abstract}

Building on the ``drPACS'' contribution at ADASS XX of a simple Unix 
pipeline infrastructure, we implemented a pipeline toolkit using the package 
MIRIAD to combine Interferometric and Single Dish data (MIS). This was 
prompted by our observations made with the Combined Array For Research 
in Millimeter-wave Astronomy (CARMA) interferometer of the star-forming 
region NGC 1333, a large survey highlighting the new 23-element and 
singledish observing modes. The project consists of 20 CARMA datasets 
each containing interferometric as well as simultaneously obtained single 
dish data, for 3 molecular spectral lines and continuum, in 527 different 
pointings, covering an area of about 8 by 11 arcminutes. A small group of 
collaborators then shared this toolkit and their parameters via CVS, and 
scripts were developed to ensure uniform data reduction across the group. 
The pipeline was run end-to-end each night as new observations were 
obtained, producing maps that contained all the data to date. We will 
show examples of the scripts and data products. This approach could 
serve as a model for repeated calibration and mapping of large mixed-mode 
correlation datasets from ALMA. 

\end{abstract}

\section{Introduction}

In a previous paper \citep{Teuben2010} a very simple and
easy to use Unix pipeline toolkit was introduced, dubbed
``drPACS''. The intent of
this toolkit was to provide an infrastructure that made it not
only easy to define a set of steps to reduce a dataset, but also
to apply these steps to many similar datasets. A set of parameters
would then control the pipeline. This could be applied to
observational data as well as theoretical data, and has
a very low buy-in cost (cf. \citet{Mandel2001}) and works
on pretty much any out-of-the-box (Unix) workstation.  The simplicity
of the pipeline also encourages the user to experiment with
parameter settings and, in what can often be a very complex 
flow of data morphing, get a better feeling for the robustness of
the results.

\section{The CARMA-23 Data}

In January 2011, CARMA commissioned its new ``CARMA-23'' mode which
correlates all baselines from the 23-element array using the new CARMA spectral
correlator. The correlator was commissioned in 2009 for 15-element 8-GHz
bandwidth operations, and new FPGA programming was later implemented to
allow 23-element operation with 4-GHz bandwidth.  CARMA is a heterogeneous
array consisting of six 10.4-m, nine 6.1-m, and eight 3.5-m antennas.
The heterogeneous nature of the array and 253 baselines allow for a
large dynamic range in sensitivity to spatial scales and excellent
imaging fidelity.

Our group obtained wide-field mosaic data of the NGC 1333 star-forming
region in the spectral lines N2H+(1-0), HCN(1-0), HCO+(1-0) and in
$\lambda$=3mm continuum.  CARMA's Python observing interface is flexible
enough that only a small change was required to simultaneously obtain the
``zero-spacing'' data by standard ON-OFF position-switching as part of
the mosaicking process.  These data were then used to recover spatial
frequency information resolved out by the interferometer.

\section{Workflow}

The foundation provided by MIS allowed us to establish a fairly simple
workflow for new data as they were obtained, enabling the team to quickly
assess the situation before the next observing run.

\begin{enumerate}
\item Download new data from CARMA Archive (essentially wget). Each dataset
  is between 200 and 900 MB.
\item Inspect and Flag bad data
 \begin{itemize}
  \item Two people separately inspected the data, and flagged any considered
    bad. They then compared their flag choices and agreed on final flags
    for that day's data.
  \item The final flagging selection was save in a MIRIAD-friendly flagging 
    definition file and checked into CVS.  All cross-correlation flags apply 
    also to auto-correlation data.
  \item Reasons for flagging were posted along with any other data 
    reduction notes to the project wiki.
 \end{itemize}
\item Derive standard interferometric calibrations (passband, phase, amplitude).
\item Update uv-coverage and integration time maps.  These were used to
modify mosaic starting point for next run of observing script in order
to keep these quantities as uniform across the map.
\item Make interferometric and singledish maps for current data.
\item Make interferometric and singledish maps using all data so far.

\end{enumerate}

\section{Example Scripts}

This procedure was encoded in the follow way, using Unix shell scripting
language:

\begin{verbatim}
  # grab copy of the data
  piperun n1333.lis     'getdata link=1'

  # process SD
  piperun n1333sd.lis   do_uvcatSD flag=1
  piperun n1333sd.lis   do_reduceSD device=/null sleep=0
  piperun n1333sd.lis   do_mapSD device=/null
  do_mapSDmedian4 

  # process INT
  piperun n1333.lis     do_uvcat1 flag=1
  piperun n1333.lis     do_inspect1
  piperun n1333.lis     'do_cal1 calflux=12.5'
  do_mos1 

  # joint deconvolution of SD and INT
  do_jd

\end{verbatim}


\section{Example Results}

Once all data for the project were obtained, the interferometric and
singledish data were combined in a joint deconvolution technique 
(see e.g. \citet{Stanimirovic2002})   
to create maps that contain all spatial scales.  Figure \ref{panelfig}
shows the HCO+, HCN, and N2H+ integrated intensity 
maps resulting from the joint deconvolution.

At the start of the pipeline the raw data amounted to about 
16 GB, growing to about 60 GB when all data have been processed
into one continuum and 
3 datacubes (for each molecule) of about 700 x 900 x 150 pixels.

\acknowledgements We like to thank our N1333 team members
Lee Mundy, Maxime Rizzo, Demerese Salter, and Shaye Storm.

\bibliography{P122}

\begin{thebibliography}{}
\expandafter\ifx\csname natexlab\endcsname\relax\def\natexlab#1{#1}\fi
\expandafter\ifx\csname url\endcsname\relax
  \def\url#1{\texttt{#1}}\fi
\expandafter\ifx\csname urlprefix\endcsname\relax\def\urlprefix{URL }\fi
\providecommand{\eprint}[2][]{\url{#2}}

\bibitem[{{Mandel} et~al.(2001){Mandel}, {Murray}, \& {Roll}}]{Mandel2001}
{Mandel}, E., {Murray}, S.~S., \& {Roll}, J.~B. 2001, in Astronomical Data
  Analysis Software and Systems X, edited by {F.~R.~Harnden Jr., F.~A.~Primini,
  \& H.~E.~Payne}, vol. 238 of Astronomical Society of the Pacific Conference
  Series, 225

\bibitem[{{Stanimirovic} et~al.(2002){Stanimirovic}, {Altschuler}, {Goldsmith},
  \& {Salter}}]{Stanimirovic2002}
{Stanimirovic}, S., {Altschuler}, D., {Goldsmith}, P., \& {Salter}, C. (eds.)
  2002, {Single-Dish Radio Astronomy: Techniques and Applications}, vol. 278 of
  Astronomical Society of the Pacific Conference Series

\bibitem[{Teuben(2011)}]{Teuben2010}
Teuben, P. 2011, in ADASS XX, edited by I.~N. Evans, A.~Accomazzi, D.~J. Mink,
  \& A.~H. Rots (San Francisco: ASP), vol. 442 of ASP Conf. Ser., 533

\end{thebibliography}

\articlefigurefour
{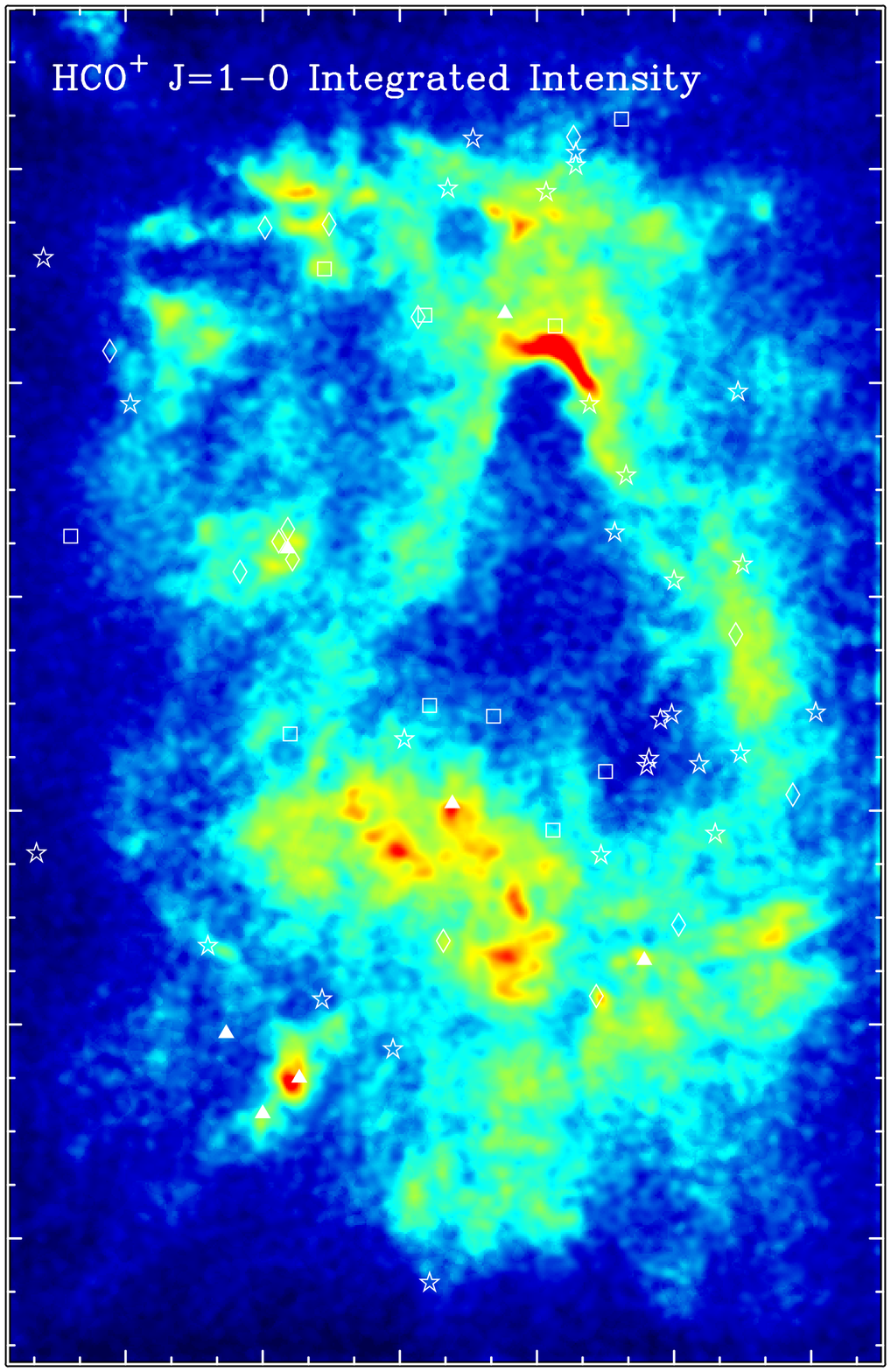}{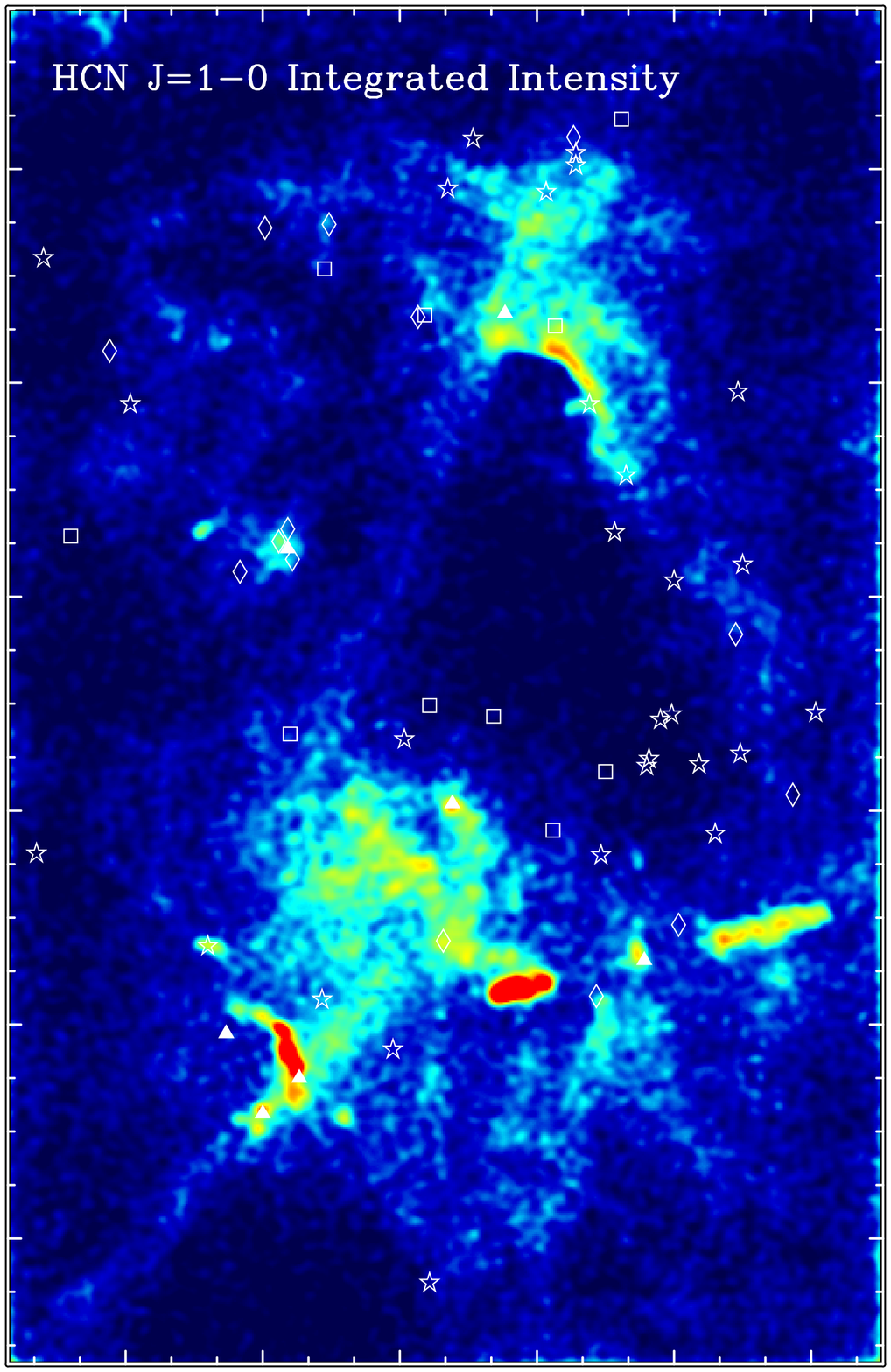}
{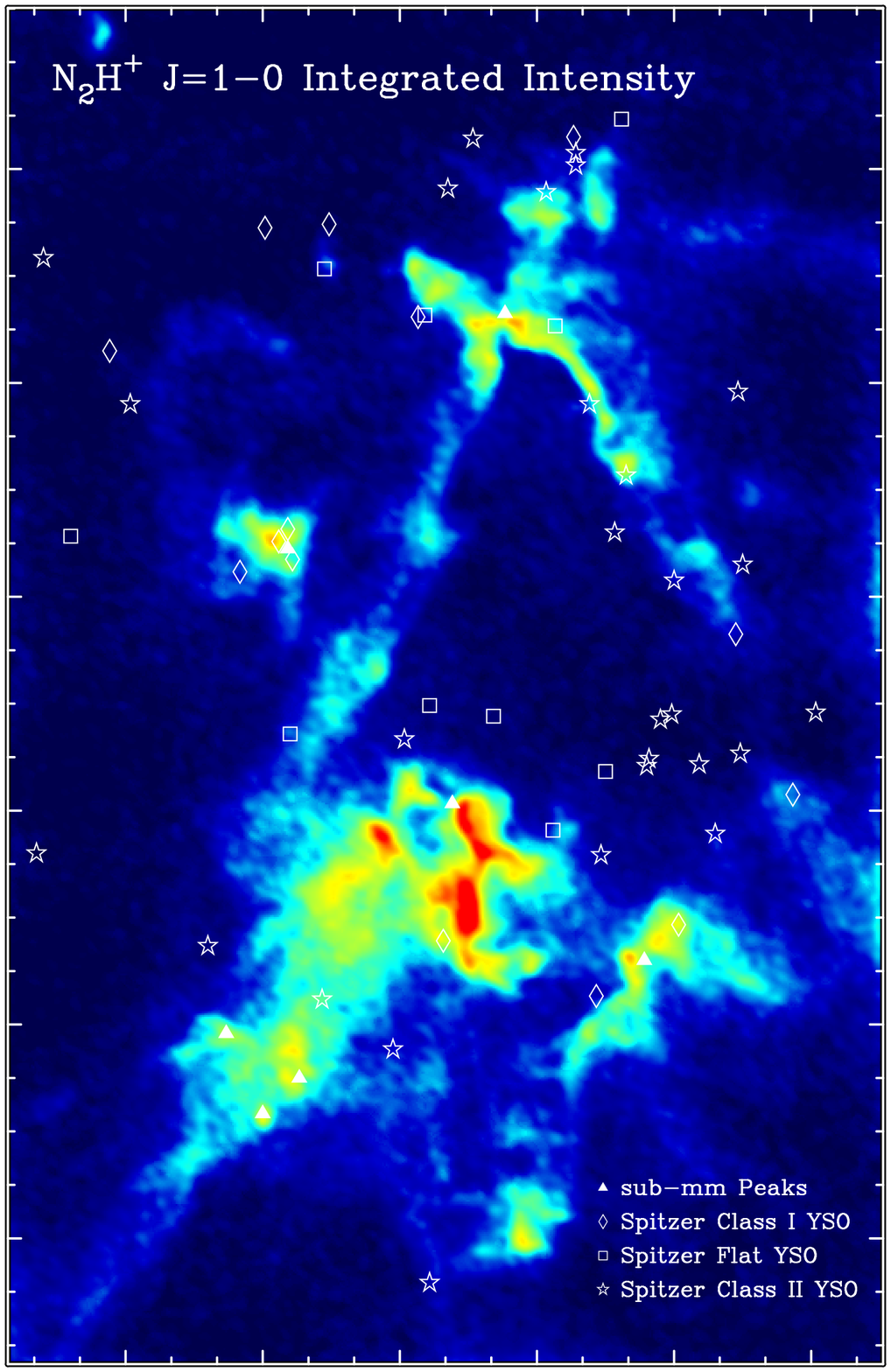}{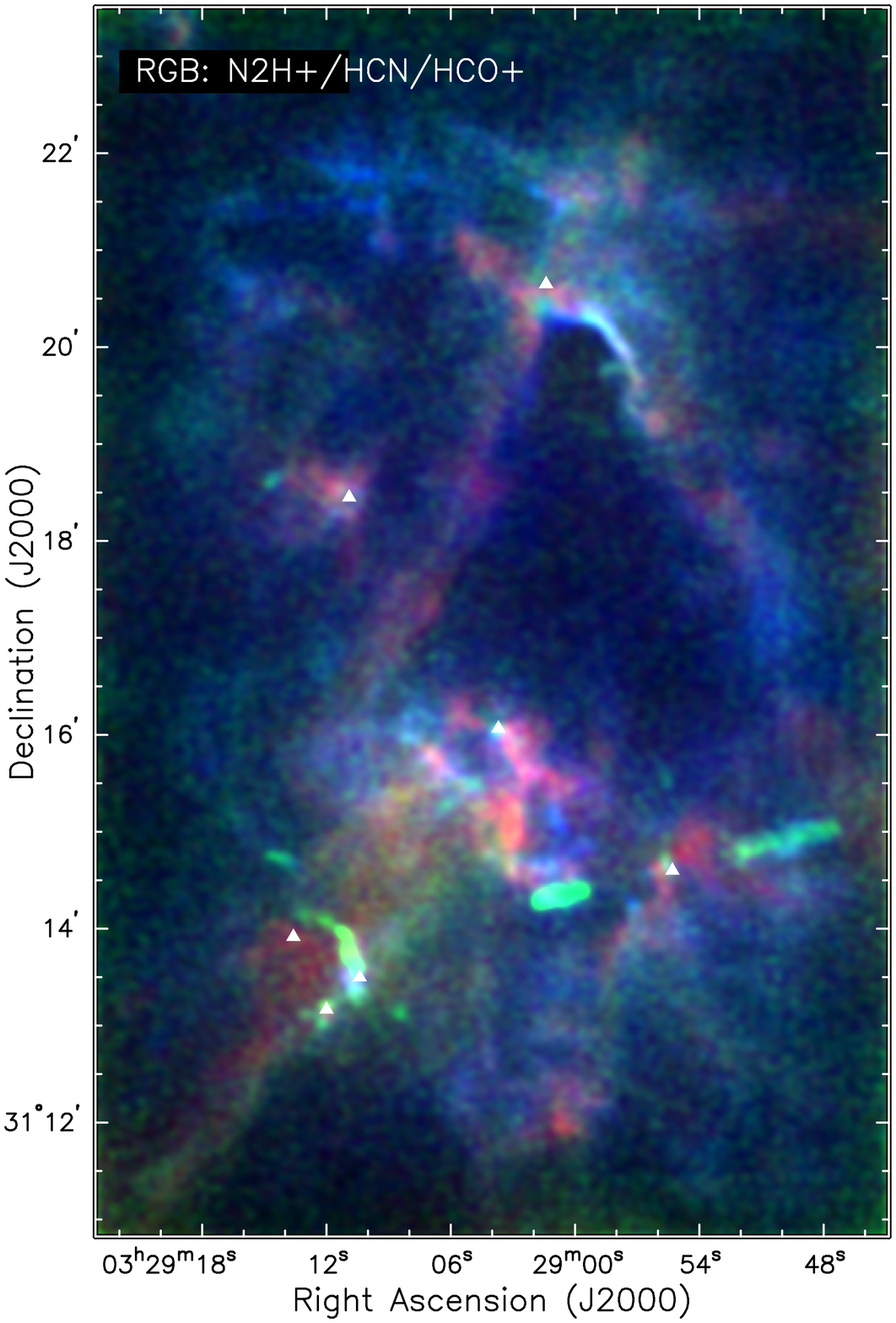}
{panelfig}{Integrated intensity maps in the three molecular lines covered 
by the CARMA survey.  Top left: HCO+(1-0);
top right: HCN(1-0); bottom left: N2H+(1-0);
bottom right: Combined RGB image of the three molecular line maps, showing
the features highlighted by the different tracers. For instance, the
protostellar outflows from IRAS2 and IRAS4 are evident in HCN (green).}

\end{document}